\documentclass[conference]{IEEEtran}

\usepackage{amssymb}

\newcommand{\A}{{\mathcal{A}}}

\newcommand{\CC}{{\mathcal{C}}}

\newcommand{\FF}{{\mathcal{F}}}

\newcommand{\HH}{{\mathcal{H}}}
\newcommand{\I}{{\mathcal{I}}}

\newcommand{\SSS}{{\mathcal{S}}}
\newcommand{\T}{{\mathcal{T}}}

\newcommand{\C}{{\mathbb{C}}}
\newcommand{\F}{{\mathbb{F}}}

\newcommand{\Z}{{\mathbb{Z}}}
\newcommand{\zerob}{{\mathbf 0}}
\newcommand{\ab}{{\mathbf a}}
\newcommand{\bb}{{\mathbf b}}

\newcommand{\fb}{{\mathbf f}}

\newcommand{\mb}{{\mathbf m}}

\renewcommand{\sb}{{\mathbf s}}
\newcommand{\tb}{{\mathbf t}}

\newcommand{\wb}{{\mathbf w}}
\newcommand{\xb}{{\mathbf x}}
\newcommand{\yb}{{\mathbf y}}
\newcommand{\zb}{{\mathbf z}}

\newcommand{\Fb}{{\mathbf F}}

\newcommand{\Mb}{{\mathbf M}}

\newcommand{\Wb}{{\mathbf W}}
\newcommand{\Xb}{{\mathbf X}}
\newcommand{\Yb}{{\mathbf Y}}
\newcommand{\Zb}{{\mathbf Z}}

\newcommand{\Bf}{{\mathfrak{B}}}

\newcommand{\ie}{{\em i.e., }}
\newcommand{\eg}{{\em e.g., }}

\newcommand{\half}{\frac{1}{2}}

\newcommand{\openbox}{\leavevmode
     \hbox to.77778em{%
     \hfil\vrule
     \vbox to.675em{\hrule width.6em\vfil\hrule}%
     \vrule\hfil}}
\newcommand{\qed}{\hspace*{1cm}\hspace*{\fill}\openbox}

\begin{document}

\title{MacWilliams Identities for Codes on Graphs}

\author{G. David Forney, Jr. \\
Laboratory for Information and Decision Systems\\
Massachusetts Institute of Technology\\
Cambridge, MA 02139\\
Email: forneyd@comcast.net}

\maketitle

\begin{abstract}
The MacWilliams identity for linear time-invariant convolutional codes that has recently been found by Gluesing-Luerssen and Schneider is proved concisely, and generalized to arbitrary group codes on graphs.  A similar development yields a short, transparent proof of the dual sum-product update rule.
\end{abstract}

\section{Introduction}

Finding a MacWilliams-type identity for convolutional codes is a problem of long standing.  Recently Gluesing-Luerssen and Schneider (GLS) have formulated  \cite{GLS08} and proved  \cite{GLS09} such an identity involving the (Hamming) weight adjacency matrix (WAM) of a linear time-invariant convolutional code over a finite field and the WAM of its orthogonal code.

The purpose of this note is to provide a concise group-theoretic proof of this identity, and to generalize it to arbitrary group codes defined on graphs.  We use first the general duality result that, given a ``normal" graphical realization of a group code $\CC$, the dual (orthogonal) code $\CC^\perp$ is realized by the dual graph, in which the ``constraint code" corresponding to each node is replaced by its orthogonal code \cite{F01}.  A more or less standard development (following \cite{F98}), using the Poisson summation formula, then proves an appropriate MacWilliams identity between the complete or Hamming WAM of a constraint code and the complete or Hamming WAM of its dual.  

In the special case of a state-space (trellis) realization of a linear time-invariant convolutional code over a finite field, all constraint codes are identical, and our result reduces to the GLS result.  Our formulation generalizes the GLS result to arbitrary group codes defined on graphs;  \eg linear time-varying convolutional codes, linear tail-biting codes, or trellis codes over finite abelian groups.

We use a similar argument to provide a concise and transparent proof of the dual sum-product update rule stated in \cite{F01}.

\section{Codes, Realizations and Graphical Models}

We follow the development and notation of \cite{F01}.

Let $\{A_k, k \in \I_\A\}$ be a set of \emph{symbol variables} $A_k$ indexed by a discrete index set $\I_\A$, where each $A_k$ is a finite abelian group.  
We will mostly consider symbol variables $A_k$ that are vector spaces over a finite field $\F$, but all of our results and proofs generalize to arbitrary finite abelian groups.

A \emph{group code} $\CC$ is a subgroup of the Cartesian-product group $\A = \Pi_{k \in \I_\A} A_k$.  If $\A$ is actually a  vector space over a finite field $\F$, then a \emph{linear code} $\CC$ is a subspace of $\A$.  From now on, all codes will be assumed to be group or linear codes.

A \emph{generalized state realization} of a code $\CC \subseteq \A$ is defined by a set of \emph{state variables} $\{S_j, j \in \I_S\}$, and a set of \emph{constraint codes} $\{\CC_i, i \in \I_\CC\}$, where $\I_S$ and $\I_\CC$ are two further discrete index sets.  Each state variable $S_j$ is a finite group, or in the linear case a vector space over $\F$.  Each constraint code $\CC_i$ is a group or linear code involving certain subsets of the symbol and state variables.  The \emph{full behavior} of the realization is the set $\Bf = (\ab, \sb)$ of all configurations of symbol variables $\ab \in \A$ and state variables $\sb \in \SSS = \Pi_{j \in \I_S} S_j$ such that all constraints are satisfied.  The \emph{code} generated by the realization is the projection $\CC = \Bf_{|\A}$ of $\Bf$ onto $\A$;  \ie the set of all symbol configurations $\ab \in \A$ that appear in some $(\ab, \sb) \in \Bf$.

For example, in a \emph{conventional state realization} of a linear code $\CC$ over a finite field $\F$, the symbol index set $\I_\A$ is a conventional  discrete time axis, namely the set of integers $\Z$, or a subinterval of $\Z$.  The state index set $\I_S$ may be thought of as the set of times that occur \emph{between} consecutive pairs of times in $\I_\A$, and the state time preceding symbol time $k \in \I_\A$ is conventionally also denoted by $k \in \I_S$.  The constraint codes $\{\CC_k, k \in I_\A\}$ are linear codes indexed by the symbol index set $\I_\A$, and specify the set of all valid $(s_k, a_k, s_{k+1})$ transitions;  \ie for each $k \in \I_\A$, $\CC_k$ is a subspace of the Cartesian product vector space $S_k \times A_k \times S_{k+1}$.
The full behavior $\Bf$ of the realization is the set of all symbol/state trajectories $(\ab, \sb)$ such that $(s_k, a_k, s_{k+1})$ is a valid transition in $\CC_k$ for all $k \in \I_\A$.  The code $\CC$ generated by the realization is the set of all symbol trajectories $\ab$ that appear  in some $(\ab, \sb) \in \Bf$.

A \emph{normal realization} is defined as a generalized state realization in which every symbol variable is involved  in precisely one constraint code, and every state variable is involved in precisely two constraint codes.  Thus a conventional state realization is normal.  It is shown in \cite{F01} that any generalized state realization may be straightforwardly converted to a normal realization by introducing replication constraints, without essentially increasing the complexity of the realization.

A normal realization has a natural graphical model, in which each constraint code $\CC_i$ corresponds to a vertex, each state variable $S_j$ (which by definition is involved in two constraints) corresponds to an edge connecting the two corresponding constraint vertices, and each symbol variable $A_k$  (which by definition is involved in one constraint) corresponds to a leaf or ``half-edge" connected to the corresponding constraint vertex.  

For example, Figure 1 shows the graph corresponding to a conventional state realization, which is a simple chain graph.  Here vertices are represented by square boxes, and the ``half-edges" corresponding to symbol variables are represented by special ``dongle" symbols.

\begin{figure}[!t]
\setlength{\unitlength}{5pt}
\centering
\begin{picture}(50,12)(-2, 1)
\multiput(0,5)(12,0){4}{\line(1,0){7}}
\multiput(9.5,7.5)(12,0){3}{\line(0,1){3}}
\multiput(8,10.5)(12,0){3}{\line(1,0){3}}
\put(-3,5){\ldots}
\put(3,6){$S_k$}
\put(13,6){$S_{k+1}$}
\put(25,6){$S_{k+2}$}
\put(37,6){$S_{k+3}$}
\put(44,5){\ldots}
\put(8,11.5){$A_k$}
\put(19,11.5){$A_{k+1}$}
\put(31,11.5){$A_{k+2}$}
\put(7,2.5){\framebox(5,5){$\CC_k$}}
\put(19,2.5){\framebox(5,5){$\CC_{k+1}$}}
\put(31,2.5){\framebox(5,5){$\CC_{k+2}$}}
\end{picture}
\caption{Graph of a conventional state realization.}
\label{Fig1}
\end{figure}
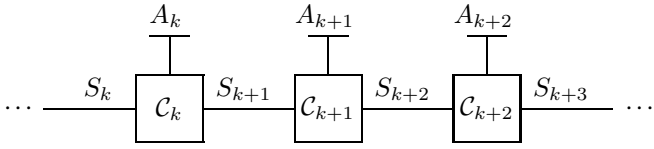

\section{Dual Normal Realizations}

The central duality result of \cite{F01} is the following:  given a normal realization of a code $\CC$, the dual normal realization generates the dual code $\CC^\perp$.  For simplicity of exposition, we will explain this result only for the case where $\CC$ is a linear code over a finite field $\F$, but it holds also in the group case;  see \cite{F01}.  In the linear case, the dual code $\CC^\perp$ is the usual orthogonal code to $\CC$ under the usual symbolwise inner product.

We have seen that a normal realization for $\CC$ is defined by a set of symbol variables $\{A_k, k \in \I_\A\}$, a set of state variables $\{S_j, j \in \I_S\}$, and a set of constraint codes $\{\CC_i, i \in \I_\CC\}$, where each symbol variable is involved in one constraint code, and each state variable is involved in two constraint codes.

The definition of a dual normal realization is slightly simpler in the case of a linear code $\CC$ over the binary field $\F_2$ than in the general case, so we discuss the binary case first.  Then the \emph{dual normal realization} is defined by the same sets of symbol and state variables, and by the set of orthogonal constraint codes $\{\CC_i^\perp, i \in \I_\CC\}$, each involving the same variables as in the primal realization.  The graph of the dual realization is thus the same as the graph of the primal realization, except that each constraint code $\CC_i$ is replaced by its orthogonal code $\CC_i^\perp$.

\vspace{1ex}
\noindent
\textbf{Example 1}.  Consider the rate-1/2 binary linear time-invariant convolutional code $\CC$ generated by the degree-2 generators $(1 + D^2, 1 + D + D^2)$, in standard $D$-transform notation.  In other words, $\CC$ is the set of all output sequences of the single-input, two-output linear time-invariant system over $\F_2$ whose impulse response is $(11, 01, 11, 00, \ldots)$.  This system has a conventional four-state realization as in Figure 1 in which each symbol variable $A_k$ may be taken as $(\F_2)^2$, each state variable $S_k$ may also be taken as $(\F_2)^2$, and each constraint code $\CC_k$ is the $(6, 3)$ binary linear block code generated by the three generators
$$
\begin{array}{c|c|c}
00 & 11 & 10; \\
10 & 01 & 01; \\
01 & 11 & 00,
\end{array}
$$
which represent the three nontrivial (state, symbol, next-state) transitions in the impulse response of the system.  The orthogonal code $\CC_k^\perp$ may easily be seen to be  the $(6, 3)$ binary linear block code generated by the three generators
$$
\begin{array}{c|c|c}
00 & 11 & 01; \\
01 & 10 & 10; \\
10 & 11 & 00,
\end{array}
$$
which represent the three nontrivial (state, symbol, next-state) transitions in the impulse response of a system with impulse response $(11, 10, 11, 00, \ldots)$, or $(1 + D + D^2, 1 + D^2)$ in $D$-transform notation.  This is indeed the generator of the orthogonal convolutional code $\CC^\perp$ under the symbolwise definition of the inner product that we are using here.  (For the more usual sequencewise definition of the inner product, we need to take the time-reversal of $\CC^\perp$,\footnote{The symbolwise inner product of two sequences $\ab, \bb \in \A$ is $\sum_{k} a_k b_k$, and that of $\ab$ and a shift of $\bb$ by $j$ time units is $\sum_{k} a_k b_{k-j}$.  The product of the corresponding $D$-transforms $a(D) = \sum_k a_kD^k$ and $b(D^{-1}) = \sum_k b_kD^{-k}$ is $\sum_j (\sum_{k} a_k b_{k-j})D^j$, so $\ab$ is orthogonal to all shifts of $\bb$ if and only if $a(D)b(D^{-1}) = 0$, or equivalently if $a(D)\tilde{b}(D) = 0$, where $\tilde{b}(D)$ is the $D$-transform of the time-reversed sequence $\tilde{\bb} = \{b_{-k}, k \in \I_\A\}$.} which in this case is again the code generated by $(1 + D + D^2, 1 + D^2)$.\qed\vspace{1ex}

For a linear code $\CC$ over a nonbinary field $\F$, one further trick (originally introduced by Mittelholzer \cite{M95} to dualize conventional state realizations over groups) is needed to define the dual normal realization:  namely, in terms of the graph of the realization, insert a sign inverter in the middle of every edge.  In other words, invert the sign of each state variable $S_k$ in one of the two constraint codes in which it is involved.  This is illustrated in Figure 2 for a conventional state realization.

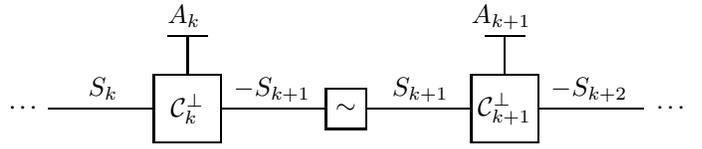
\begin{figure}[!t]
\setlength{\unitlength}{5pt}
\centering
\begin{picture}(50,12)(-2, 1)
\multiput(-1,5)(24,0){2}{\line(1,0){8}}
\multiput(12,5)(24,0){2}{\line(1,0){8}}
\multiput(9.5,7.5)(24,0){2}{\line(0,1){3}}
\multiput(8,10.5)(24,0){2}{\line(1,0){3}}
\put(-4,5){\ldots}
\put(2,6){$S_k$}
\put(13,6){$-S_{k+1}$}
\put(25,6){$S_{k+1}$}
\put(37,6){$-S_{k+2}$}
%\put(49,6){$S_{k+2}$}
%\put(61,6){$-S_{k+3}$}
\put(45,5){\ldots}
\put(8,11.5){$A_k$}
\put(31,11.5){$A_{k+1}$}
%\put(55,11.5){$A_{k+2}$}
\put(7,2.5){\framebox(5,5){$\CC_k^\perp$}}
\put(31,2.5){\framebox(5,5){$\CC_{k+1}^\perp$}}
%\put(55,2.5){\framebox(5,5){$\CC_{k+2}^\perp$}}
\put(20,3.5){\framebox(3,3){$\sim$}}
%\put(44,3.5){\framebox(3,3){$\sim$}}
\end{picture}

\caption{Graph of dual of a conventional state realization, with sign inverter.}
\label{Fig2}
\end{figure}

\vspace{1ex}
\noindent
\textbf{Example 2} (\emph{cf.\ }\cite{GLS08, GLS09}).  Consider the rate-2/3 linear time-invariant convolutional code $\CC$ over $\F_3$ with $g_1(D) = (1 + D^2, 2 + D, 0)$ and $g_2(D) = (1, 0, 2)$.  In other words, $\CC$ is the set of all output sequences of the two-input, three-output linear time-invariant system over $\F_3$ whose impulse responses are $(120, 010, 100, 000, \ldots)$ and $(102, 000, \ldots).$  This system has a conventional nine-state realization as in Figure 1 in which each symbol variable $A_k$ may be taken as $(\F_3)^3$, each state variable $S_k$ may be taken as $(\F_3)^2$, and each constraint code $\CC_k$ is the $(7, 4)$ ternary linear block code generated by the four generators  
$$
\begin{array}{c|c|c}
00 & 120 & 10; \\
10 & 010 & 01; \\
01 & 100 & 00; \\
00 & 102 & 00,
\end{array}
$$
which represent the four nontrivial ($s_k, a_k, s_{k+1}$) transitions in the two impulse responses of the system.  The orthogonal code $\CC_k^\perp$ is  the $(7, 3)$ ternary linear block code generated by the three generators
$$
\begin{array}{c|c|c}
00 & 010 & 12; \\
21 & 202 & 11; \\
22 & 111 & 00, 
\end{array}
$$
\pagebreak
which represent the three nontrivial ($s'_k, a'_k, -s'_{k+1}$) transitions in the impulse response of a conventional state realization of a single-input, three-output linear system over $\F_3$, with sign inverters as in Figure 2, whose impulse response is $(010, 202, 111, 000, \ldots)$, or $(2D + D^2, 1 + D^2, 2D + D^2)$ in $D$-transform notation.  (Note the unconventional basis of the dual state space.)  This is indeed the generator of the orthogonal convolutional code $\CC^\perp$ under our symbolwise definition of the inner product.  (For the more usual sequence-wise definition of the inner product, we need to take the time-reversal of $\CC^\perp$, which in this case is the code generated by $(1 + 2D, 1 + D^2, 1 + 2D)$.)\qed

\section{MacWilliams Identities}

Given these duality results, various MacWilliams-type identities may be obtained in a more or less standard manner.  We follow the development in \cite{F98}.

Every finite abelian group $\T$ is a direct product of cyclic groups.  In particular, every finite field $\F$ has $q = p^m$ elements for some prime $p$ and is isomorphic as an additive group to $(\Z_p)^m$, and every vector space over a finite field $\F_{p^m}$ of dimension $d$ is isomorphic to $(\Z_p)^{md}$.  Thus, for some integer $n$, we may take $\T = (\Z_p)^n$, the set of $n$-tuples of elements of $\Z_p$.

Given a complex-valued function $\{x: \T \to \C, t \mapsto x(t)\}$ defined on $\T = (\Z_p)^n$, its (Fourier) \emph{transform} is the complex-valued function $\{X: \FF \to \C, f \mapsto X(f)\}$ defined on $\FF = (\Z_p)^n$ by
$$
X(f) = \sum_\T x(t)  \omega^{f\cdot t}, \quad f \in \FF,
$$
where $\omega$ is a primitive complex $p$th root of unity, and $f\cdot t \in \Z_p$ is the ordinary dot product between the $n$-tuples $f \in (\Z_p)^n$ and $t \in (\Z_p)^n$ over $\Z_p$.  If we view $\xb = \{x(t), t \in \T\}$ as a vector indexed by $\T$, and similarly $\Xb = \{X(f), f \in \FF\}$ as a vector indexed by $\FF$, then the transform can be expressed in matrix form as
$$
\Xb = \HH \xb,
$$
where the \emph{transform matrix} is $\HH = \{\omega^{f\cdot t}, f \in \FF, t \in \T\}$.  Note that $\HH^T = \HH$, where $\HH^T$ denotes the transpose of $\HH$.

From the \emph{orthogonality relation} 
$$
\sum_\FF  \omega^{f\cdot t} = \left\{
\begin{array}{cc}
 |\FF|,  & t = 0; \\ 
 0, & t \neq 0, 
 \end{array} \right.
$$
we obtain the matrix equation 
$$\HH \HH^* = |\FF| I_{|\FF|},$$
 where $\HH^* = \{\omega^{-f\cdot t}, f \in \FF, t \in \T\}$, $|\FF| = |\T| =  p^n$, and $I_{|\FF|}$ is the $|\FF| \times |\FF|$ identity matrix.  In other words, the inverse of $\HH$ is $\HH^{-1} = |\FF|^{-1} \HH^*$.
Thus we obtain the \emph{inverse transform}
$$
\xb = \HH^{-1} \Xb = \frac{\HH^*\Xb}{|\FF|}.
$$
We say that $\xb$ and $\Xb$ are a \emph{transform pair}.

We may extend these definitions to a set of indeterminates $\zb = \{z(t), t \in \T\}$ indexed by $\T$, rather than a complex-valued function.  The transform of this set is then a set of indeterminates $\Zb = \{Z(f), f \in \FF\}$ indexed by $\FF$, where
$$
\Zb = \HH \zb.
$$
Again, we have the inverse transform relationship
$$
\zb = \HH^{-1} \Zb = \frac{\HH^*\Zb}{|\FF|},
$$
and we say that $\zb$ and $\Zb$ are a transform pair.   

For example, if $\T = \Z_2$, then $Z(0) = z(0) + z(1)$ and $Z(1) = z(0) - z(1)$;  similarly, $z(0) = \half(Z(0) + Z(1))$, and $z(1) = \half(Z(0) -Z(1))$.

Now let us consider weight enumerators, initially for the case of a conventional state realization over a finite field $\F$ as in Figure 1.  We define the \emph{complete weight adjacency matrix} (CWAM) of each constraint code $\CC_k \subseteq S_k \times A_k \times S_{k+1}$ as follows.  

If $A_k = \F^n$, then define the complete weight enumerator of the $n$-tuple $\ab = (a_1, \ldots, a_n \in \F^n)$ as the product $w(\ab) = \Pi_{1 \le i \le n} w(a_i)$, where $\wb = \{w(a), a \in \F\}$ is a set of indeterminates indexed by $\F$.  Then the CWAM of $\CC_k$ is the matrix $\Lambda(\wb) = \{\Lambda(s_k, s_{k+1})(\wb), (s_k, s_{k+1}) \in S_k \times S_{k+1}\}$ defined by 
$$
\Lambda(s_k, s_{k+1})(\wb) = \sum_{\CC(s_k, s_{k+1})} w(\ab_k),
$$
where $\CC(s_k, s_{k+1}) = \{\ab_k \mid (s_k, \ab_k, s_{k+1}) \in \CC_k\}$.  
Thus each entry $\Lambda(s_k, s_{k+1})(\wb)$ is a homogeneous integer polynomial of degree $n$ in the $|\F|$ indeterminates $\wb = \{w(a), a \in \F\}$.

Now let $\yb = \{y(s_k), s_k \in S_k\}$ and $\zb = \{z(s_{k+1}), s_{k+1} \in S_{k+1}\}$ be sets of indeterminates indexed by the state variables $S_k$ and $S_{k+1}$, respectively, and define a \emph{generating function} $g_{\Lambda(\wb)}(\yb, \zb)$, a polynomial in the sets of indeterminates $\yb$ and $\zb$, as follows:
\begin{eqnarray*}
g_{\Lambda(\wb)}(\yb, \zb) & = & \yb^T \Lambda(\wb) \zb  \\ & = &
\sum_{S_k \times S_{k+1}} y(s_k)\Lambda(s_k, s_{k+1})(\wb)z(s_{k+1}).
\end{eqnarray*}
From the definition $\CC(s_k, s_{k+1}) = \{\ab_k \mid (s_k, \ab_k, s_{k+1}) \in \CC_k\}$, it follows that
$$
g_{\Lambda(\wb)}(\yb, \zb) = 
\sum_{(s_k, \ab_k, s_{k+1}) \in \CC_k} y(s_k)w(\ab_k)z(s_{k+1}).
$$

\vspace{1ex}
\noindent
\textbf{Example 1 (cont.)}.  The constraint code of Example 1 has the eight codewords 
$00|00|00, 00|11|10, 10|01|01,$ $10|10|11, 01|11|00, 01|00|10, 11|10|01, 11|01|11,$
corresponding to the eight possible (state, symbol, next-state) transitions.  Writing $\{w_0, w_1\}$ instead of $\{w(0), w(1)\}$, we see that we may write the CWAM of this constraint code in matrix form as 
$$
\Lambda(\wb) \quad  = \quad
\begin{array}{c|c|c|c|c|}
s_k/s_{k+1} & 00 & 10 & 01 & 11 \\
\hline
00 & w_0^2 & w_1^2 & 0 & 0 \\
\hline
10 & 0 & 0 & w_0w_1 & w_0w_1 \\
\hline
01 & w_1^2 & w_0^2 & 0 & 0 \\
\hline
11 & 0 & 0  & w_0w_1 & w_0w_1 \\
\hline
\end{array}
$$
Equivalently, its generating function $g_{\Lambda(\wb)}(\yb, \zb)$ is
\begin{eqnarray*}
g_{\Lambda(\wb)}(\yb, \zb) & = & w_0^2 (y_{00} z_{00} + y_{01} z_{10}) + w_1^2 (y_{00} z_{10} + y_{10} z_{00}) \\ & + & w_0 w_1 (y_{01} z_{01} + y_{01} z_{11} + y_{11} z_{01} + y_{11} z_{11}).
\end{eqnarray*}
\qed

The key duality relation for MacWilliams identities is the \emph{Poisson summation formula}, which says that ``the sum of a function over a linear space is equal to the sum of the Fourier transform of the function over the dual space" \cite{CS88}.  For our case, this formula may be stated as follows:

\vspace{1ex}
\noindent
\textbf{Poisson summation formula}.  Let $\xb$ and $\Xb$ be a transform pair defined on $\T = (\Z_p)^n$ and $\FF = (\Z_p)^n$, respectively, and let $\CC$ and $\CC^\perp$ be orthogonal subgroups of $\T$ and $\FF$, respectively.  Then
$$
\sum_{t \in \CC} x(t) = \frac{1}{|\CC^\perp|} \sum_{f \in \CC^\perp} X(f).
$$

Now, applying this formula to the equation above for $g_{\Lambda(\wb)}(\yb, \zb)$, we obtain
\begin{eqnarray*}
g_{\Lambda(\wb)}(\yb, \zb) & = & \sum_{\CC_k} y(s_k)w(\ab_k)z(s_{k+1}) \\ & = & \frac{1}{|\CC_k^\perp|} \sum_{\CC_k^\perp} Y(\hat{s}_k)W(\hat{\ab}_k)Z(-\hat{s}_{k+1}). % = \frac{g_{\hat{\Lambda}(\Wb)}(\Yb, \Zb)}{|\CC_k^\perp|},
\end{eqnarray*}
Here we use the fact that the transform of a product is the product of their transforms, where $\Yb = \HH_y \yb$, $\Wb = \HH_w \wb$, and $\Zb = \HH_z \zb$.  Note that $\Wb$ is itself a product transform.  Also, since the elements of $\HH_z$ are $\HH_z(s_{k+1}, -\hat{s}_{k+1}) = \omega^{-s_{k+1}\cdot\hat{s}_{k+1}}$, the matrix $\HH_z$ is the conjugate of the usual transform matrix over $S_{k+1}$.

If we define the CWAM $\hat{\Lambda}(\Wb)$ of $\CC_k^\perp$ and its generating function $g_{\hat{\Lambda}(\Wb)}(\Yb, \Zb)$ similarly to the analogous quantities for $\CC_k$, then we obtain
$$
g_{\hat{\Lambda}(\Wb)}(\Yb, \Zb) = \sum_{\CC_k^\perp} Y(\hat{s}_k)W(\hat{\ab}_k)Z(-\hat{s}_{k+1}).
$$
Using inverse transforms, we thus obtain 
\begin{eqnarray*}
g_{\hat{\Lambda}(\Wb)}(\Yb, \Zb) & = &   |\CC_k^\perp| g_{\Lambda(\wb)}(\yb, \zb) \\ & = &
|\CC_k^\perp| g_{\Lambda(\HH_w^{-1} \Wb)}(\HH_y^{-1}\Yb, \HH_z^{-1}\Zb).
\end{eqnarray*}
This MacWilliams identity shows how the generating function for the CWAM $\hat{\Lambda}(\Wb)$ of $\CC_k^\perp$ may be obtained from that for $\CC_k$, or \emph{vice versa}.  

Alternatively, since 
$$g_{\hat{\Lambda}(\Wb)}(\Yb, \Zb) = \Yb^T \hat{\Lambda}(\Wb) \Zb$$ and 
$$
g_{\Lambda(\wb)}(\yb, \zb) = \yb^T \Lambda(\wb) \zb = \Yb^T \HH_y^{-1} \Lambda(\HH_w^{-1} \Wb) \HH_z^{-1} \Zb,
$$
we may simply write
$$
\hat{\Lambda}(\Wb) = |\CC_k^\perp|   \HH_y^{-1} \Lambda(\HH_w^{-1} \Wb) \HH_z^{-1} ,
$$
a MacWilliams identity that shows how
the CWAM of $\CC_k^\perp$ may be obtained from the CWAM of $\CC_k$.

\vspace{1ex}
\noindent
\textbf{Example 1 (cont.)}.  Given the CWAM $\Lambda(\wb)$ of the constraint code $\CC_k$ of Example 1, the CWAM  $\hat{\Lambda}(\Wb)$ of the orthogonal constraint code $\CC_k^\perp$ is given by the matrix equation at the top of the next page, where we have substituted the dual indeterminates $W_0$ and $W_1$ for $w_0 + w_1$ and $w_0 - w_1$. \qed
\vspace{1ex}

\begin{figure*}[!t]
$$
\half
 \left[ \begin{array}{crrr}
1 & 1 & 1 & 1 \\
1 & -1 & 1 & -1 \\
1 & 1 & -1 & -1 \\
1 & -1 & -1 & 1 \\
\end{array}  \right]
 \left[ \begin{array}{cccc}
w_0^2 & w_1^2 & 0 & 0 \\
0 & 0 & w_0w_1 & w_0w_1 \\
w_1^2 & w_0^2 & 0 & 0 \\
0 & 0  & w_0w_1 & w_0w_1 \\
\end{array}  \right]
 \left[ \begin{array}{crrr}
1 & 1 & 1 & 1 \\
1 & -1 & 1 & -1 \\
1 & 1 & -1 & -1 \\
1 & -1 & -1 & 1 \\
\end{array}  \right]
=
 \left[ \begin{array}{cccc}
W_0^2 & 0 & W_1^2 & 0 \\
W_1^2 & 0 & W_0^2 & 0 \\
0 & W_0W_1 & 0 & W_0W_1 \\
0 & W_0W_1 & 0 & W_0W_1 \\
\end{array}  \right]
$$
\hrule
\end{figure*}

The Hamming weight adjacency matrix (HWAM) $\Lambda_H$ of a constraint code $\CC_k$ is obtained by substituting $1$ for $w(0)$ and $w$ for each $w(a), a \neq 0$.  Thus each element $\Lambda_H(s_k, s_{k+1})(w)$ becomes a polynomial of degree $n$ in the single indeterminate $w$.  The dual indeterminates become $W(0) = 1 + (|\F| - 1)w$ and $W(a) = 1 - w, a \neq 0$, which scale to $1$ and $W = (1-w)/(1 + (|\F|-1)w)$, respectively.  Substituting in the above MacWilliams-type identities for CWAMs, we obtain MacWilliams-type identities for HWAMs.  This yields the main result of \cite{GLS08, GLS09}.\footnote{The MacWilliams identity of \cite{GLS08, GLS09} is stated in terms of the HWAM for a minimal realization of a linear time-invariant convolutional code $\CC$ in controller canonical form, and the HWAM of \emph{some} minimal encoder for the orthogonal code $\CC^\perp$.  Our results apply to the CWAM or HWAM of any state realization, and the CWAM or HWAM of its dual realization, because in our development, by constraint code duality, the basis of the dual state space representation is fixed as soon as the basis of the primal state space is fixed.}

\vspace{1ex}
\noindent
\textbf{Example 2 (cont.)}.  For a worked-out example of  the HWAM  $\hat{\Lambda}(W)$ of the orthogonal code $\CC_k^\perp$ to the constraint code of Example 2, see \cite{GLS09}. \qed
\vspace{1ex}

Although our development has focussed on conventional state realizations of linear time-invariant convolutional codes, it may be straightforwardly extended to obtain MacWilliams identities for any generalized state realization of any finite abelian group code defined on an arbitrary graph, because constraint code duality holds in the general case.  

\section{Dualizing the Sum-Product Update Rule}

Another duality result in \cite{F01} is a general method for dualizing the sum-product update rule, which among other things yields the ``tanh rule" of APP decoding.  The approach of this paper yields a cleaner derivation of this result. 

Again, for simplicity we restrict attention to conventional state realizations, in which each constraint code $\CC_k$ specifies the state transitions in $S_k \times A_k \times S_{k+1}$ that can possibly occur.  Let the (right-going) \emph{message} be any real- or complex-valued function $\mb_k = \{m_k(s_k), s_k \in S_k\}$ of the state variable $S_k$, and let $\fb_k = \{f_k(a_k), a_k \in A_k\}$ be any real-or complex-valued \emph{weight function} of the symbol variable $A_k$.  Then the \emph{sum-product update rule} associated with constraint code $\CC_k$ is
$$
m_{k+1}(s_{k+1}) = \sum_{\CC_k(s_{k+1})} m_k(s_k)f_k(a_k),
$$
where $\CC_k(s_{k+1}) = \{(s_k, a_k) \in S_k \times A_k \mid (s_k, a_k, s_{k+1}) \in \CC_k\}$.  In other words, if we define a set of indeterminates $\xb = \{x(s_{k+1}), s_{k+1} \in S_{k+1}\}$, then $m_{k+1}(s_{k+1})$ is the coefficient of $x(s_{k+1})$ in the homogeneous degree-1 multivariate generating function $g_{k+1}(\xb)$ defined by
$$
g_{k+1}(\xb) = \mb_{k+1}^T \xb = \sum_{s_{k+1}} m_{k+1}(s_{k+1})x(s_{k+1}).
$$
From the definition of $\CC_k(s_{k+1})$, it follows that
$$
g_{k+1}(\xb) = \sum_{\CC_k} m_k(s_k)f_k(a_k)x(s_{k+1}).
$$
Using the Poisson summation formula, we obtain
\begin{eqnarray*}
g_{k+1}(\xb) & = &
\sum_{\CC_k} m_k(s_k)f_k(a_k)x(s_{k+1}) \\ & = & \frac{1}{|\CC_k^\perp|} \sum_{\CC_k^\perp} M_k(\hat{s}_k)F_k(\hat{a}_k)X(-\hat{s}_{k+1}) \\ & = & \frac{\hat{g}_{k+1}(\Xb)}{|\CC_k^\perp|},
\end{eqnarray*}
where we again use the fact that the transform of a product is the product of their transforms, and define transformed functions or indeterminates by corresponding capitalized functions or indeterminates.  The left side of this equation is the generating function $g_{k+1}(\xb)$ of the message $\{m_{k+1}\}$, and the right side is  (up to scale) the generating function $\hat{g}_{k+1}(\Xb)$ of the message $\Mb_{k+1}$ obtained by performing the sum-product update algorithm for $\CC_{k}^\perp$ upon the message $\Mb_k$ and the weight function $\Fb_k$.  Moreover, the messages $\mb_{k+1}$ and $\Mb_{k+1}$ form a transform pair.

Consequently, we have the following recipe for performing the sum-product update rule for $\CC_k$:
\begin{enumerate}
\item Transform the incoming messages $\mb_k$ and $\fb_k$ to $\Mb_k$ and $\Fb_k$;
\item Perform the sum-product update rule for $\CC_k^\perp$ to generate an output message $\Mb_{k+1}$;
\item Inverse transform $\Mb_{k+1}$ to obtain the message $\mb_{k+1}$, up to the scale factor $|\CC_k^\perp|$.
\end{enumerate}
Since the complexity of performing the sum-product update rule for $\CC_k$ is proportional to $|\CC_k|$, this dual computation may be attractive if $|\CC_k^\perp| < |\CC_k|$.
\pagebreak

\vspace{1ex}
\noindent
\textbf{Example 3 (``tanh rule")}.  Let $S_k, A_k$ and $S_{k+1}$ be binary variables taking values in $\F_2$, and let $\CC_k$ be the $(3, 2)$ single-parity-check code consisting of the four codewords $(000, 011, 101, 110)$;  then $\CC_k^\perp$ is the $(3, 1)$ repetition code consisting of the two codewords $(000, 111)$.  Let the incoming message and weight function be $\mb_k = (m_0, m_1)$ and  $\fb_k = (f_0, f_1)$;  then the transformed message and weight function are $\Mb_k = (M_0 = m_0 + m_1, M_1 = m_0 - m_1)$ and $\Fb_k = (F_0 = f_0 + f_1, F_1 = f_0 - f_1)$.  Using two multiplications, the sum-product update equation then produces the message $\Mb_{k+1} = (M_{k+1}(0) = (m_0 + m_1)(f_0 + f_1), M_{k+1}(1) = (m_0 - m_1)(f_0 - f_1))$. Thus, up to scale, the message $\mb_{k+1}$ is 
\begin{eqnarray*}
m_{k+1}(0) & = & M_{k+1}(0) + M_{k+1}(1) \propto m_0f_0 + m_1 f_1;  \\
m_{k+1}(1) & = & M_{k+1}(0) - M_{k+1}(1) \propto m_0 f_1 + m_1 f_0; 
\end{eqnarray*}
which is evidently the message that would have been computed by a direct computation of the sum-product update rule for $\CC_k$, which requires four multiplications. \qed
\vspace{1ex}

Again, although our development has focussed on a constraint code of a conventional linear state realization, it may be straightforwardly extended to obtain a dual sum-product update rule for an arbitrary constraint code over any finite abelian group.

\section*{Acknowledgment}  
For comments on an earlier version of this paper, I am grateful to H. Gluesing-Luerssen.

\end{document}